\documentclass{webofc}
\usepackage[varg]{txfonts}
\usepackage{graphicx,amsmath,amssymb,wrapfig}
\usepackage{slashed}
\begin{document}
\title{Tomography and gravitational radii for hadrons\\
       by three-dimensional structure functions}
\author{\firstname{S.} \lastname{Kumano}\inst{1,2}
\and\firstname{Qin-Tao} \lastname{Song}\inst{1}
\and\firstname{O. V.} \lastname{Teryaev}\inst{3}}
\institute{KEK Theory Center, Institute of Particle and Nuclear Studies, KEK\\
\ \ \            and Department of Particle and Nuclear Physics,
Graduate University for Advanced Studies \\
\ \ \  (SOKENDAI),       Ooho 1-1, Tsukuba, Ibaraki, 305-0801, Japan\\ 
\and
J-PARC Branch, KEK Theory Center,
     Institute of Particle and Nuclear Studies, KEK\\
\ \ \ and Theory Group, Particle and Nuclear Physics Division, J-PARC Center,\\
\ \ \ 203-1, Shirakata, Tokai, Ibaraki, 319-1106, Japan\\
\and
Bogoliubov Laboratory of Theoretical Physics,
            Joint Institute for Nuclear Research,\\ 
\ \ \ 141980 Dubna, Russia}

\abstract{
Three-dimensional tomography of hadrons can be investigated by
generalized parton distributions (GPDs), 
transverse-momentum-dependent parton distributions (TMDs), and
generalized distribution amplitudes (GDAs).
The GDA studies had been only theoretical for a long time because
there was no experimental measurement until recently, whereas
the GPDs and TMDs have been investigated extensively
by deeply virtual Compton scattering and semi-inclusive deep
inelastic scattering.
Here, we report our studies to determine pion GDAs from recent
KEKB measurements on the differential cross section of
$\gamma^* \gamma \to \pi^0 \pi^0$.
Since an exotic-hadron pair can be produced in the final state,
the GDAs can be used also for probing internal structure of exotic
hadron candidates in future. The other important feature of the GDAs is
that the GDAs contain information on form factors of the energy-momentum 
tensor for quarks and gluons, so that gravitational form factors and radii 
can be calculated from the determined GDAs. We show the mass (energy) and 
the mechanical (pressure, shear force) form factors and radii for the pion.
Our analysis should be the first attempt for obtaining gravitational
form factors and radii of hadrons by analysis of actual experimental
measurements. We believe that a new field of gravitational physics is created
from the microscopic level in terms of elementary quarks and gluons.
}

\maketitle

\section{Introduction}
\label{intro}

Inclusive lepton deep inelastic scattering (DIS) has been investigated 
since 1970's and it is described by structure functions 
and parton distribution functions (PDFs) expressed by
the Bjorken-scaling variable $x$. 
Since it is the longitudinal momentum fraction for a parton in a hadron,
the inclusive DIS probes the one-dimensional structure of hadrons.
However, it became necessary to understand three-dimensional (3D)
structure of hadrons for precisely describing exclusive and 
semi-inclusive reactions and for finding the origin of nucleon spin 
including contributions from partonic orbital angular momenta.
As the 3D structure functions, generalized parton distributions (GPDs) 
and transverse-momentum-dependent parton distributions (TMDs)
have been investigated both theoretically and experimentally.
They are measured by the deeply virtual Compton scattering 
(DVCS) and semi-inclusive deep inelastic lepton scattering.
There is another type of 3D structure functions called 
generalized distribution amplitudes (GDAs), which can be
investigated by the $s$-$t$ crossed process to the DVCS
($\gamma^* h \to \gamma h$),
namely the two-photon process $\gamma^* \gamma \to h \bar h$.

\begin{wrapfigure}[10]{r}{0.52\textwidth}
   \vspace{-0.55cm}
   \begin{center}
     \includegraphics[width=7.1cm]{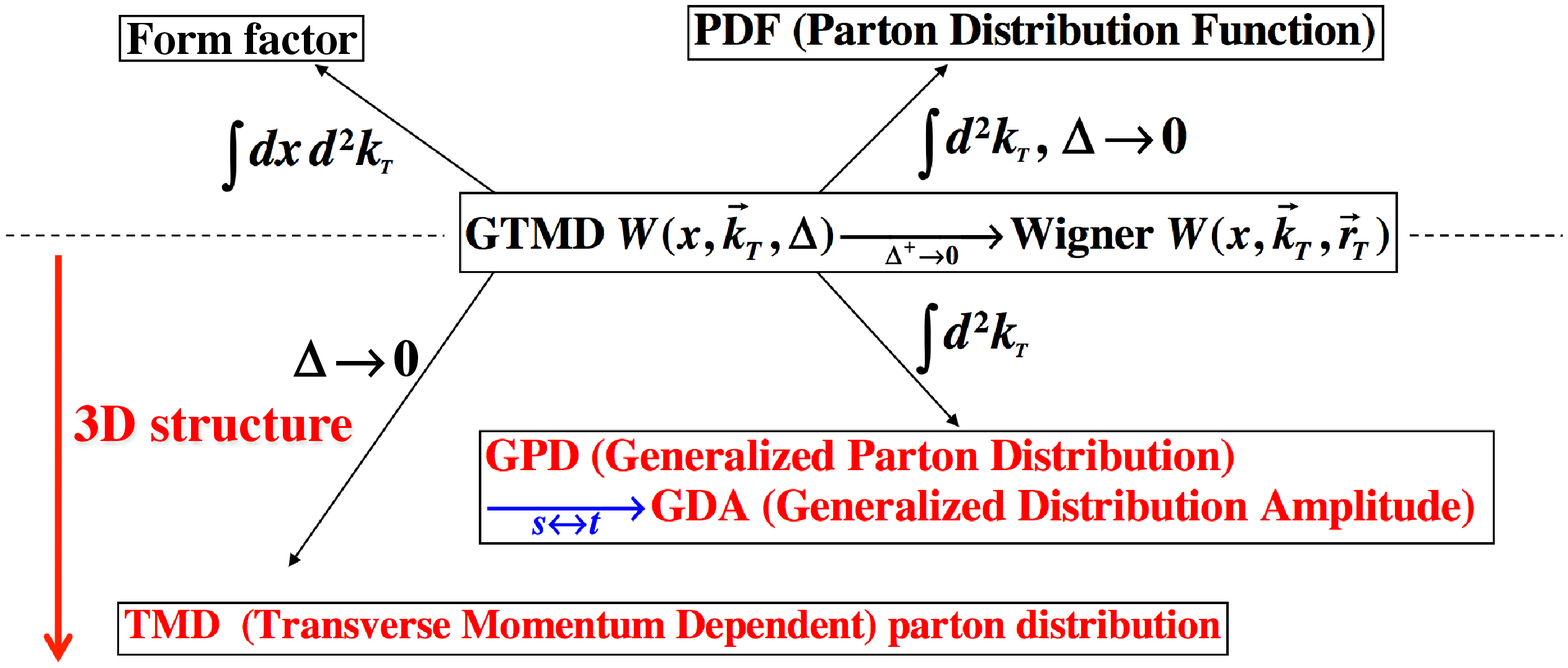}
   \end{center}
\vspace{-0.80cm}
\caption{3D structure functions and Wigner distributions.}
\label{fig:wigner-3d-sfs}
\vspace{-0.7cm}
\end{wrapfigure}

The situation is illustrated in Fig.\,\ref{fig:wigner-3d-sfs}.
The nucleonic PDFs and form factors have been investigated for a long time,
and recent studies focus on the 3D structure functions, GPDs, TMDs, and GDAs.
These functions are obtained by integrating the generating functions,
generalized transverse-momentum-dependent parton distributions (GTMDs)
or Wigner distributions. Although there are much experimental progress
on the GPDs and TMDs in the last several years, 
there was no experimental information on the GDAs until recently.
However, the Belle collaboration reported the cross-section data
on the two-photon process $\gamma^* \gamma \to \pi^0 \pi^0$ in 2016
\cite{Masuda:2015yoh}, so that it became possible to extract 
the pion GDAs from their data \cite{gdas-kst-2017}.

The determination of the GDAs is valuable in studying 3D tomography
of hadrons for finding the origin of nucleon spin, because
the GPDs and GDAs are related by the $s$-$t$ crossing 
and because both distributions are obtained from common double
distributions by using the Radon transform as explained in Sect.\,\ref{gdas}.
Such tomography studies could be used also for probing internal structure 
of exotic hadron candidates because an unstable hadron pair could be produced
in the two-photon process \cite{gdas-kk-2014}, whereas unstable hadrons 
cannot be used as fixed targets in measuring the GPDs and TMDs.
There is another important advantage in studying the GPDs and GDAs
for probing gravitational source by the energy-momentum tensor
of quarks and gluons.

The electric charge and magnetic form factors of the nucleons
are measured in electron scattering and their radii are determined
from them. The charge radius of the pion is measured as
$0.672 \pm 0.008$ fm.
The charged particles in the pion, namely quarks, contribute
to the charge form factor and the radius.
In the same way, it is interesting to measure the gravitational mass 
distributions and radii for the pion or any hadron.
Here, both quarks and gluons contribute to the gravitational distributions.
It is not measured in direct scattering experiment like
the electron scattering because of the ultra-weak nature 
of gravitational interactions.
However, there is a way to access them by the 3D structure functions,
such as the GDAs, because they contain the factors of the energy-momentum tensor
for quarks and gluons \cite{gdas-kst-2017}. 
We know the the energy-momentum tensor is the source of gravity.

We discuss determination of the pion GDAs and gravitational form factors
by analyzing the Belle measurements in this report \cite{gdas-kst-2017}. 
First, the GDAs are introduced in Sect.~\ref{gdas},
and the cross section of the two-photon process $\gamma^*\gamma\to \pi^0\pi^0$
is explained with the GDAs in Sect.~\ref{cross-section}.
Analysis results are shown in Sect.\,\ref{results},
and our studies are summarized in Sect.~\ref{summary}

\section{Generalized distribution amplitudes
         and gravitational form factors}
\label{gdas}

\begin{figure}[t!]
\vspace{-0.30cm}
\begin{minipage}{\textwidth}
\begin{tabular}{lcl}
\begin{minipage}[c]{0.45\textwidth}
    \includegraphics[width=5.0cm]{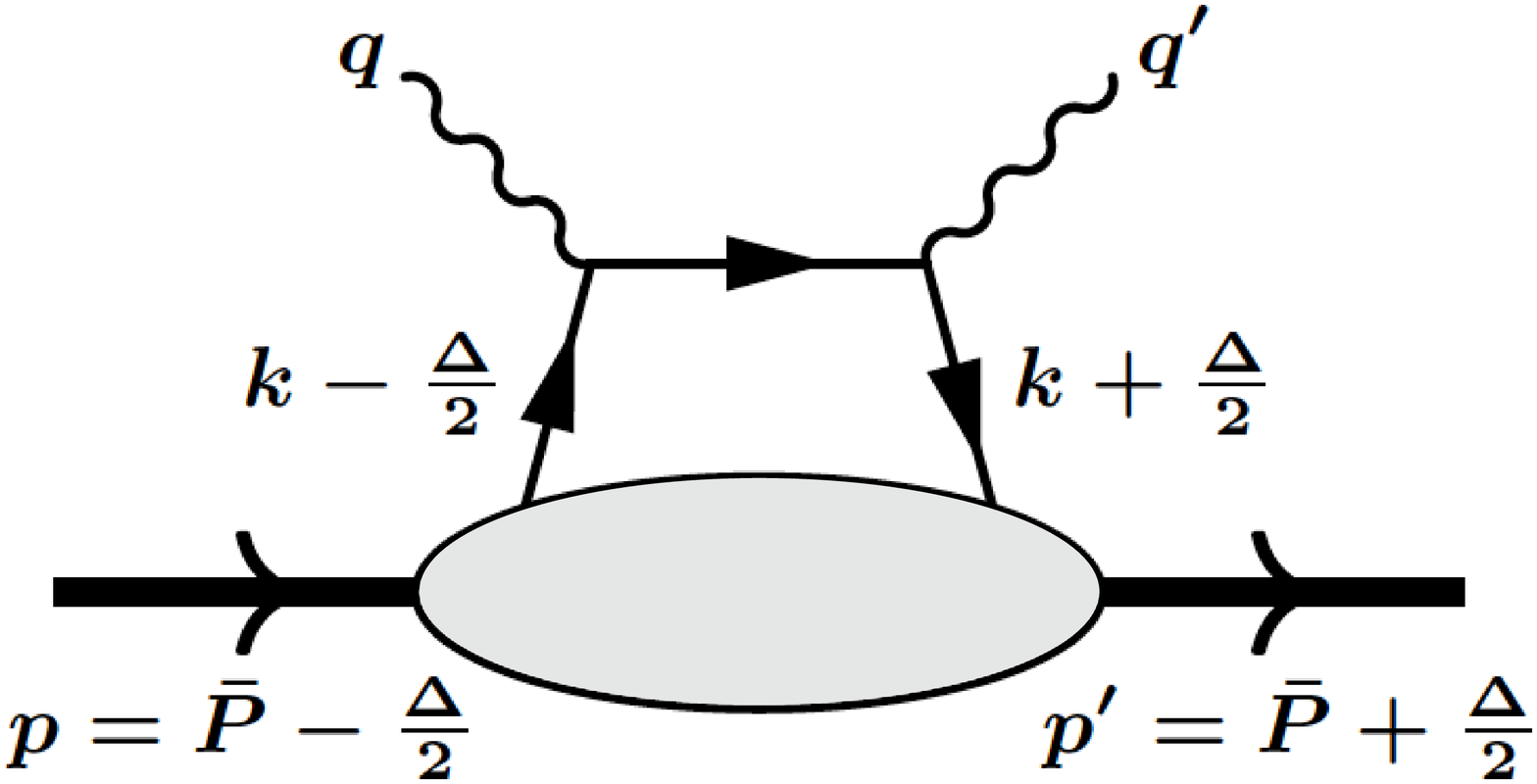}
\end{minipage} 
\hspace{0.50cm}
\begin{minipage}[c]{0.45\textwidth}
    \includegraphics[width=4.2cm]{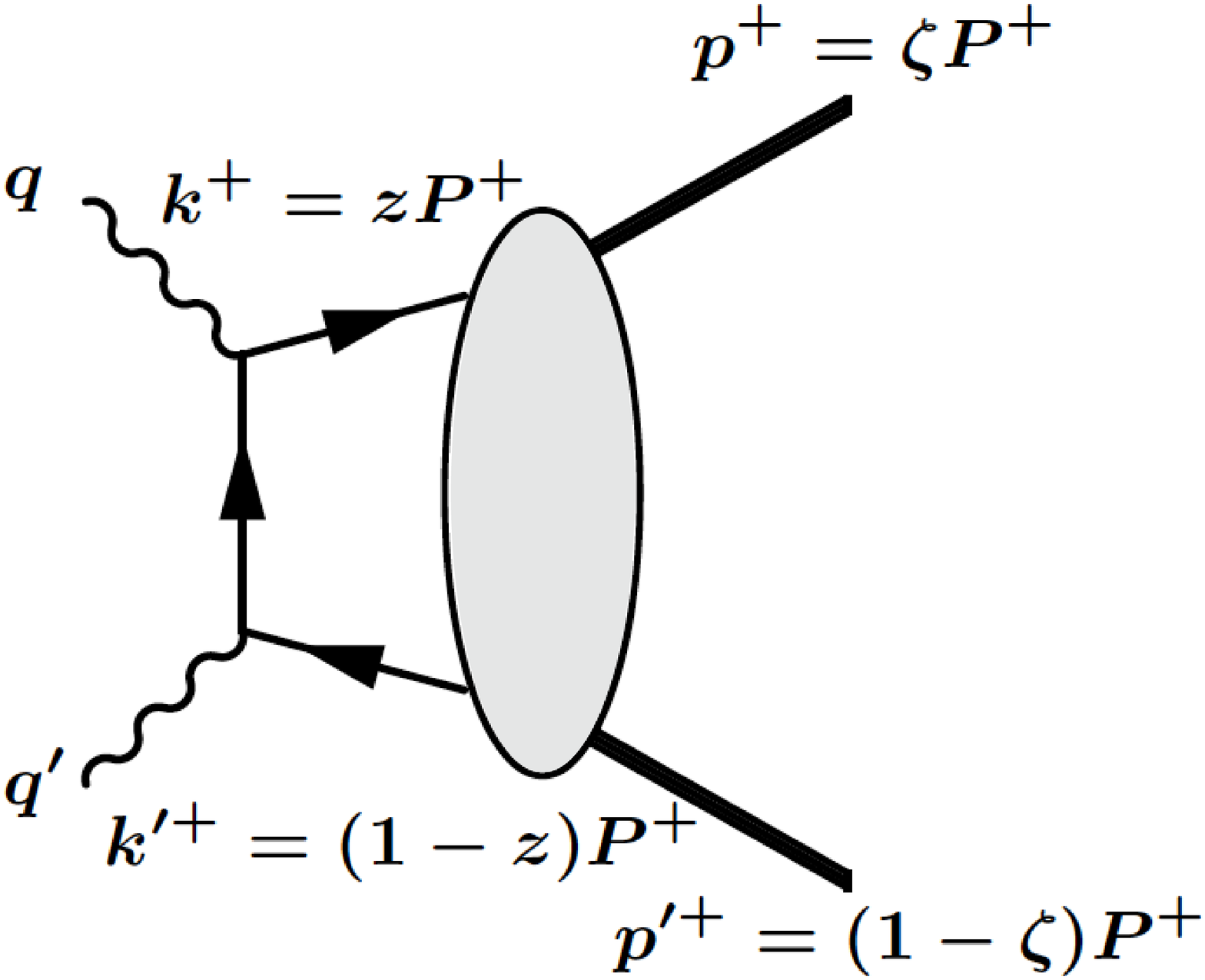}
\end{minipage}
\end{tabular}
\end{minipage}
\ \vspace{-0.00cm}\hspace{2.9cm}
$(a)$ \hspace{5.5cm} $(b)$
\vspace{-0.2cm}
\caption{\ $(a)$ Virtual Compton process for GPDs
\ \ \ \ \ \ \ \ \ \ 
$(b)$ Two-photon process for GDAs.}
\label{fig:gpd-fig}
\vspace{-0.5cm}
\end{figure}

We explain the GDAs in comparison with the GPDs, which have been 
studied extensively. First, the GPDs can be experimentally studied 
by the DVCS $\gamma^* h \to \gamma h$
in Fig.\,\ref{fig:gpd-fig}$(a)$
if $Q^2 = -q^2$ is large enough to satisfy the factorization
that the process is described by the hard perturbative 
QCD part and the soft GPD one. 
Here, $q$ is the initial virtual-photon momentum.
The $s$-$t$ crossed process of the DVCS is the two-photo process 
to produce a hadron pair $\gamma^*\gamma\to h \bar h$ as shown 
in Fig.\,\ref{fig:gpd-fig}$(b)$. 
It is also factorized into the hard part and the soft GDA one 
if $Q^2$ is large enough.

The pion GPDs $H_q^{\, \pi^0}$ are defined by off-forward matrix elements
of quark and gluon operators with a lightcone separation, and
the GDAs $\Phi_q^{\, \pi^0 \pi^0}$ are defined by the same operator
between the vacuum and the hadron pair as \cite{gpds-gdas-rev,gdas}
\begin{align}
H_q^{\, \pi^0} (x,\xi,t)
& = \int\frac{d y^-}{4\pi} \, e^{i x \bar P^+ y^-}
 \left< \pi^0 (p') \left| 
\overline{q}(-y/2) \gamma^+ q(y/2) 
 \right| \pi^0 (p) \right> \Big |_{y^+ = \vec y_\perp =0}, 
\nonumber \\
\Phi_q^{\, \pi^0 \pi^0} (z,\zeta,W^2) 
& = \int \frac{d y^-}{2\pi}\, e^{i (2z-1)\, P^+ y^- /2}
  \langle \, \pi^0 (p) \, \pi^0 (p') \, | \, 
 \overline{q}(-y/2) \gamma^+ q(y/2) 
  \, | \, 0 \, \rangle \Big |_{y^+=\vec y_\perp =0} \, .
\label{eqn:gpd-gda-pi}
\end{align}
These GPDs and GDAs are defined for quarks, and similar expressions
exist also for gluons. Here, link operators for the color gauge invariance 
are not explicitly written for simplicity.
The PDFs are given by the forward limit 
of the GPDs: $q^{\pi^0}(x) = H_q^{\pi^0} (x,\xi=0,t=0)$ for quarks
at $x>0$ (antiquark distributions at $x<0$).
Using the initial and final pion (photon) momenta $p$ and $p'$
($q$ and $q'$), we define average momenta ($\bar P$, $\bar q$)
and momentum transfer $\Delta$ as
\begin{align}
\bar P = \frac{p+p'}{2} , \ \ 
\bar q = \frac{q+q'}{2} , \ \ 
\Delta = p'-p = q-q' .
\end{align}
Then, the Bjorken variable $x$, the skewdness parameter $\xi$,
and the momentum-transfer squared $t$ are given by
\begin{align}
x = \frac{Q^2}{2p \cdot q} , \ \ \ \ 
\xi = \frac{\bar Q^2}{2 \bar P \cdot \bar q} , \ \ \ \ 
t = \Delta^2 , 
\end{align}
where $Q^2=-q^2$ and $\bar Q^2=-\bar q^2$.
The DVCS process is factorized,
if the kinematical condition $Q^2 \gg |t|,\ \Lambda_{\text{QCD}}^2$
is satisfied, to express it in terms of the GPDs
in Fig.\,\ref{fig:gpd-fig}$(a)$. 
Here, $\Lambda_{\text{QCD}}$ is the QCD scale parameter.

The variables of the GDAs are the momentum fractions
$z$ and $\zeta$ in Fig.\,\ref{fig:gpd-fig}$(b)$
and the invariant-mass squared $W^2$,
and they are defined by
\begin{align}
z = \frac{k \cdot q'}{P \cdot q'} = \frac{k^+}{P^+} , \ \ \ 
\zeta = \frac{p \cdot q'}{P \cdot q'}
      = \frac{p^+}{P^+} = \frac{1+\beta \cos\theta}{2} , \ \ \ 
W^2 = (p+p')^2 = (q+q')^2 = s ,
\end{align}
where $\beta$ is the pion velocity given by
$\beta =|\vec p \,|/p^0 = \sqrt{1-4m_\pi^2/W^2}$,
and the scattering angle is $\theta$ in the center-of-mass frame
of final pions.
The two-photon process is factorized and expressed by the GDAs 
if the condition $Q^2 \gg W^2,\ \Lambda_{\text{QCD}}^2$ is satisfied.
The GPDs and GDAs are related with each other by the $s$-$t$ crossing as 
\begin{align}
\Phi_q^{\,\pi^0 \pi^0} (z',\zeta,W^2) 
\longleftrightarrow
H_q^{\,\pi^0} \left ( x=\frac{1-2z'}{1-2\zeta},
            \, \xi=\frac{1}{1-2\zeta}, \, t=W^2 \right ) .
\label{eqn:gda-gpd-relation}
\end{align}
However, the physical regions of the GDAs 
($0 \le z \le 1$, $|1-2\zeta| \le 1$, $W^2 \ge 0$)
do not necessarily correspond to the physical ones of the GPDs
($|x| \le 1$, $|\xi| \le 1$, $t \le 0$):
\begin{align}
0 \le |x| < \infty, \ \ 
0 \le |\xi| < \infty, \ \ 
|x| \le |\xi| , \ \ 
t \ge 0 .
\label{eqn:gda-gpd-kinematics}
\end{align}
Therefore, the GDA studies may not be directly utilized for
clarifying the GPDs. There is a way to circumvent this issue
by using the Radon transform, which is often used in 
tomography studies in general.

Let us consider possible two-pion states in the reaction 
$\gamma^* \gamma \to \pi \pi$.
The isospin $I=1$ $\pi \pi$ state is antisymmetric under the exchange 
of the pions, whereas the $I=0$ and $I=2$ $\pi \pi$ states are symmetric.
The $C$ parity of the $\pi \pi$ state is $C=(-1)^{L+S}=(-1)^L=1$ with $S=0$
because the $C$ parity of $\gamma^* \gamma$ is even. It means $L$ is even. 
Then, the Pauli principle $(-1)^L (-1)^I (-1)^S=1$ suggests that
the $\pi \pi$ isospin state is $I=0$ or 2. However, the GDAs are defined
by the vector-type nonlocal operator and the isospin of $\bar{q}q$ is 0 or 1,
so that the $\pi\pi$ states from the two photons are 
$I=0$ with $L=\text{even numbers (0,\,2,\,$\cdots$)}$.
Therefore, the GDAs in the process $\gamma^* \gamma \to \pi^0 \pi^0$
are $C$-even functions denoted with $(+)$:
\begin{align}
\Phi_q^{\pi^0 \pi^0} (z,\zeta,W^2) 
= \Phi^{\pi\pi (I=0)} (z,\zeta,W^2)  
= \Phi_q^{\pi\pi (+)} (z,\zeta,W^2)  .
\label{eqn:2pion0-gdas}
\end{align}
The charge-conjugation and isospin symmetries
require the relations for the GDAs:
\begin{align}
\Phi_q^{\pi^0 \pi^0} (z,\zeta,W^2) 
& = - \Phi_q^{\pi^0 \pi^0} (1-z,\zeta,W^2) 
 =   \Phi_q^{\pi^0 \pi^0} (z,1-\zeta,W^2) ,
\nonumber \\
\Phi_u^{\pi^0 \pi^0} (z,\zeta,W^2) 
& = \Phi_d^{\pi^0 \pi^0} (z,\zeta,W^2) ,
\label{eqn:pion-gdas-relatiohns}
\end{align}
which are considered as constraints in setting up
the parametrization of the $\pi^0$ GDAs.

The Radon transform is defined for a function $f (x)$
in $n$ dimensions as \cite{radon-book}
\begin{align}
\hat f (p, \xi) 
   = \int d^n x \, f(x) \, \delta (p - \xi \cdot x) ,
\label{eqn:readon-define}
\end{align}
\begin{wrapfigure}[7]{r}{0.35\textwidth}
   \vspace{-0.6cm}
   \begin{center}
     \includegraphics[width=4.6cm]{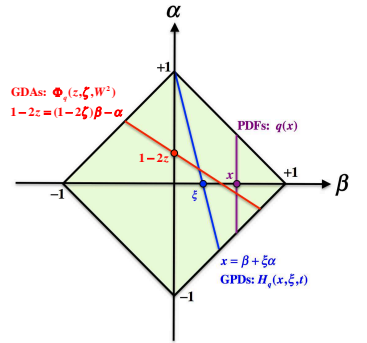}
   \end{center}
\vspace{-0.80cm}
\caption{Radon transforms for PDFs, GPDs, and GDAs.}
\label{fig:radon-trans}
\end{wrapfigure}
\noindent
where $x$ is the $n$-dimensional space coordinate
[$x$\,=\,($x_1$,\,$x_2$, $\cdots$,\,$x_n$)]
and $\xi$ is the unit vector in $n$ dimensions
[$\xi$\,=\,($\xi_1$,\,$\xi_2$, $\cdots$,\,$\xi_n$)].
We express the GPDs and GDAs by the common double distributions (DDs)
$F_q$, $G_q$, and $D_q$ with different Radon transforms given by
\cite{teryaev-2001} 
\begin{align}
H_q (x, \xi, t)  =  
\int  &  d\beta \, d\alpha \, \delta (x-\beta- \xi \alpha) 
 \left [ \, F_q (\beta,\alpha,t) + \xi G_q (\beta,\alpha,t) \, \right ] ,
\nonumber \\
\Phi_q^{h \bar h} (z,\zeta,W^2 )
 = & -2 (1-2\zeta)  \int d\beta \, d\alpha 
\delta \big( 1-2z-(1-2\zeta)\beta+\alpha \big) \,
\nonumber \\
& \ \ \ \ \ \ \ \ \ \ \ \ \ \ \ \times
F_q (1-2z,\alpha,W^2)
     -2 D_q (x/\xi,W^2) ,
\nonumber \\
q (x)  = & \int_{-1+x}^{1-x} d\alpha \,  F_q (\beta,\alpha,t=0) .
\label{eqn:dd-gdas}
\end{align}

\noindent
These integral paths are shown in Fig.\,\ref{fig:radon-trans}.
Namely, the GPDs are obtained by integrating the DDs over the slight line
$x=\beta+ \xi \alpha$, 
the PDFs are by the integral over the vertical line 
with the condition of the forward limit ($t=0$), and
the GDAs are by the Radon transform along the different line
$1-2z-(1-2\zeta)\beta+\alpha =0$.
If the DDs are determined by the GDA studies, they can be used
for finding the GPDs and vice verse. Therefore, the GDA studies
should be valuable for clarifying the 3D tomography of hadrons
and also finding the origin of nucleon spin including
orbital-angular-momentum contributions.\\
\indent
There is an important application of the GPD and GDA studies
for investigating gravitational aspects of hadrons.
The GPDs and GDAs are defined in Eq.\,(\ref{eqn:gpd-gda-pi})
with the common nonlocal operator. If the $n$-th moment
of this operator is calculated, we obtain
\begin{align}
\! \!
2(P^+/2)^{n} \! \! \int_0^1 dz \, (2z-1)^{n-1} 
  \! \!  \int\frac{d y^-}{2\pi}e^{i (2z-1) P^+ y^- /2}
\overline{q}(-y/2) \gamma^+ q(y/2) \Big |_{y^+ = \vec y_\perp =0}
 = \overline q (0) \gamma^+ \!
 \left ( i \overleftrightarrow \partial^+  \right )^{n-1} 
 q(0) .
\label{eqn:tensor-int}
\end{align} 

\begin{wrapfigure}[8]{r}{0.51\textwidth}
   \vspace{-0.4cm}
   \begin{center}
     \includegraphics[width=6.0cm]{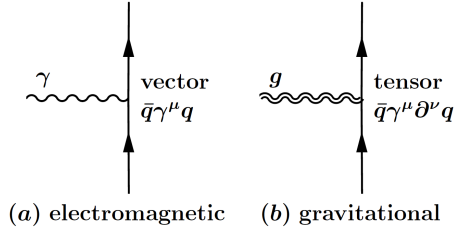}
   \end{center}
\vspace{-0.60cm}
\caption{Electromagnetic and gravitational form factors.}
\label{fig:electro-grav}
\vspace{-0.7cm}
\end{wrapfigure}
\noindent
We notice in this equation that the right-hand side is
the usual vector-type current 
$\bar q \gamma^{\,\mu} q$ for $n=1$ and the second ($n=2$) moment is
the energy-momentum tensor $\bar q \gamma^{\,\mu} i \partial^\nu q$
for quarks as shown in Fig.\,\ref{fig:electro-grav}.
Actually, the second moment is expressed as
\cite{gdas}
\begin{align}
& \int_0^1  dz \, (2z -1) \, 
\Phi_q^{\pi^0 \pi^0} (z,\,\zeta,\,W^2) 
\nonumber \\
& \ \ \ \ 
= \frac{2}{(P^+)^2} \langle \, \pi^0 (p) \, \pi^0 (p') \, | \, T_q^{++} (0) \,
       | \, 0 \, \rangle .
\label{eqn:integral-over-z}
\end{align}
In general, the quark energy-momentum tensor is defined by
$ T_q^{\,\mu\nu} (x) = \overline q (x) \, \gamma^{\,(\,\mu} 
   i \overleftrightarrow D^{\nu)} \, q (x)$,
where $D^\mu$ is the covariant derivative 
$D^\mu = \partial^{\,\mu} -ig \lambda^a A^{a,\mu}/2$
with the QCD coupling constant $g$ 
and the SU(3) Gell-Mann matrix $\lambda^a$.

The matrix element of the energy-momentum tensor is expressed
in terms of the timelike gravitational form factors $\Theta_1 (s)$ and $\Theta_2 (s)$
as
\begin{align}
\langle \, \pi^0 (p) \, \pi^0 (p') \, | \, T_q^{\mu\nu} (0) \, | \, 0 \, \rangle 
= \frac{1}{2} 
  \left [ \, \left ( s \, g^{\,\mu\nu} -P^{\,\mu} P^\nu \right ) \, \Theta_{1, q} (s)
                + \Delta^\mu \Delta^\nu \,  \Theta_{2, q} (s) \,
  \right ] ,
\label{eqn:emt-ffs-timelike-0}
\end{align}
where $P=p+p'$ and $\Delta=p'-p$.
Using Eqs.\,(\ref{eqn:integral-over-z}) and (\ref{eqn:emt-ffs-timelike-0}),
we can calculate the timelike gravitational form factors $\Theta_1 (s)$ 
and $\Theta_2 (s)$ if the GDAs are determined by analyzing experimental data on 
$\gamma^* \gamma \to \pi^0 \pi^0$.
The spacelike gravitational form factors $\Theta_1 (t)$ and $\Theta_2 (t)$
are defined by the matrix element as
\begin{align}
\langle \, \pi^0 (p') \, | \, T_q^{\mu\nu} (0) \, | \, \pi^0 (p) \, \rangle 
= \frac{1}{2} 
  \left [ \, \left ( t \, g^{\,\mu\nu} -q^{\,\mu} q^\nu \right ) \, \Theta_{1, q} (t)
                + P^{\,\mu} P^\nu \,  \Theta_{2, q} (t) \,
  \right ] ,
\label{eqn:emt-ffs-spacelike}
\end{align}
where $q=p'-p$.
The spacelike form factors are calculated by using the dispersion relation
from the timelike ones, which are obtained directly from the determined GDAs.
As explained in Sect.\,\ref{results},
the form factor $\Theta_2$ indicates the gravitational mass (energy)
distribution and $\Theta_1$ is the mechanical (pressure, shear force)
distribution.

\section{Cross section for two-photon process $\gamma^*\gamma\to \pi^0\pi^0$
        and GDAs}
\label{cross-section}

We analyze cross-section measurements for the two-photon process
$\gamma^*\gamma\to \pi^0\pi^0$ and try to determine the GDAs 
of the pion. The cross section is expressed by the matrix element
${\cal M} (\gamma^* \gamma \to \pi^0 \pi^0 )$ as
\begin{align}
d\sigma = \frac{1}{4 q\cdot q'}
\underset{\lambda, \lambda'}{\overline\sum}
| {\cal M} (\gamma^* \gamma \to \pi^0 \pi^0 ) |^2 \,
\frac{d^3 p}{(2\pi)^3 \, 2E_p}  \frac{d^3 p'}{(2\pi)^3 \, 2E_{p'}} 
(2\pi)^4 \delta^4(q+q'-p-p') ,
\label{eqn:cross-section}
\end{align}
and it is written by the hadron tensor ${\cal T}_{\mu\nu}$
and photon-polarization vectors $\epsilon^{\,\mu}(\lambda)$
and $\epsilon^\nu(\lambda')$ as
$ i {\cal M} (\gamma^* \gamma \to \pi^0 \pi^0 ) 
   = \epsilon^\mu(\lambda) \, \epsilon^\nu(\lambda') \, {\cal T}_{\mu\nu} $.
The hadron tensor is generally defined by the matrix element of 
the electromagnetic current $J_\mu ^{em}$ and then by the GDAs
for the pion, if the hard scale satisfies the factorization
condition $Q^2 \gg W^2,\ \Lambda_{\text{QCD}}^2$,
in the leading order of the running coupling constant
$\alpha_s$ and leading twist \cite{gdas-kst-2017}:
\begin{align}
\! \! \!
{\cal T}_{\mu \nu } 
 = i \! \! \int \! d^4 y \, {e^{ - iq \cdot y}} \!
\left\langle \pi^0 (p) \pi^0 (p') \! \left| 
{TJ_\mu ^{em}(y)J_\nu ^{em}(0)} \right|0 \right\rangle 
=  - g_{T}^{\, \mu \nu}{e^2} 
\sum\limits_q \frac{{e_q^2}}{2} \!
\int_0^1 \! \! {dz} \frac{{2z - 1}}{{z(1 - z)}}
\Phi_q^{\pi^0 \pi^0}(z,\zeta ,{W^2}) ,
\label{eqn:matrix}
\end{align}
where
$ g_{T}^{\, \mu \nu} = -1 $  for $\mu=\nu=1, \ 2$ and
$ g_{T}^{\, \mu \nu} = 0 $  for $\mu$, $\nu=\,$others.
The helicity amplitude $A_{i j}$ is defined as
$ A_{i j}  = \varepsilon _\mu ^{( i )}(q) \, \varepsilon _\nu ^{( j )}(q') \,
{{\cal T}^{\mu \nu }} /e^2$ ($i=-,\ 0, \ + \, ; \ j=-,\ + \, $), and
the differential cross section is expressed by the helicity amplitude
$A_{++}$:
\begin{align}
\frac{d\sigma}{d(\cos \theta)}
= \frac{\pi \alpha^2}{4(Q^2+s)}
    \sqrt{1-\frac{4m_\pi^2}{s}} \, |A_{++}|^2  , \ \ \ 
A_{++}
=\sum_q \frac{e_q^2}{2} \int^1_0 dz \frac{2z-1}{z(1-z)} 
   \Phi_q^{\pi^0 \pi^0} (z, \xi, W^2) ,
\label{eqn:cross2}
\end{align}
where the parity-conservation relation $A_{--} = A_{++}$ is used.
The gluon GDA contributes in the higher-order of $\alpha_s$,
and the terms $A_{0+}$ and $A_{0-}$ are higher-twist ones.
They are neglected in our analysis.

Because of the page limitation, the details of the GPD parametrization 
are not discussed in this article and they should be found in 
Ref.\,\cite{gdas-kst-2017}. Only the outline is explained
in the following.
First, the GPDs should satisfy the symmetry relations 
in Eq.\,(\ref{eqn:pion-gdas-relatiohns}).
The asymptotic ($Q^2 \to \infty$) $z$-dependence form is given by
$z (1-z) (2z-1)$ for the $\pi^0$ GDAs, so that the parameter $\alpha$
may be assigned to its functional form as $z^\alpha (1-z)^\alpha (2z-1)$.
Since there are S- and D-wave contributions to the $\pi^0 \pi^0$ state,
the quark GPDs are expressed as
\begin{align}
\Phi_q^{\pi ^0 \pi^0} (z, \zeta, W^2) & = 
           N_\alpha z^\alpha(1-z)^\alpha (2z-1) \,
 [\widetilde B_{10}(W^2) + \widetilde B_{12}(W^2) P_2(\cos \theta)] , 
\label{eqn:gda-parametrization}
\end{align}
where $P_2(\cos \theta)$ is the Legendre polynomial.
The S- and D-wave terms are $\widetilde B_{10}(W^2)$ 
and $\widetilde B_{12}(W^2)$, respectively, and they 
are given by the GDA continuum part and resonance 
contributions from $f_0 (500)$ and $f_2 (1270)$
\cite{teryaev-2005-f2}:
\begin{align}
\widetilde{B}_{10}(W^2) 
& = - \bigg[ \,  \left( 1+\frac{2\, m_\pi^2}{W^2} \right)
\frac{10}{9} M_{2(q)}^\pi   
F^{\,\pi}_q (W^2)
+ \sum_{f_0}
\frac{5 \, g_{f_0\pi\pi} \, \bar f_{f_0}}
    {3 \sqrt{2} \sqrt{(M^2_{f_0}-W^2)^2+\Gamma^2_{f_0} M^2_{f_0} }} \, \bigg] \, 
     e^{i \delta_0 (W)}      ,
\nonumber \\
\widetilde{B}_{12}(W^2)
& =  \left( 1 - \frac{4\, m_\pi^2}{W^2} \right) \frac{10}{9} \,
 \bigg[  \,
 M_{2(q)}^\pi
F^{\,\pi}_q (W^2)
+ \frac{g_{f_2\pi\pi} \, f_{f_2} M^2_{f_2} \beta^2 }
     {\sqrt{2} \sqrt{(M^2_{f_2}-W^2)^2+\Gamma^2_{f_2} M^2_{f_2} }} \, \bigg] \,
      e^{i\delta_2 (W)}  .
\label{eqn:B10B12-final}
\end{align}
The decay constants $\bar f_{f_0}$ and $f_{f_2}$ ($\equiv f_f$ ) have 
$Q^2$ dependence, and it is expressed as
\begin{align}
f_{f} (Q^2) = f_{f} (Q_0^2)  \,
  \left [ \frac{\alpha_s (Q^2)}{\alpha_s (Q_0^2)} 
      \right ]^{\gamma_n/\beta_0} \! \! ,  \ \ \ \ 
\gamma_n = C_F \left [ \, 1 - \frac{2}{(n+1)(n+2)} 
            + 4 \sum_{j=2}^{n+1} \frac{1}{j} \, \right ] ,
\label{eqn:ff-f02-evol-3}
\end{align}
effectively by including the scale dependence 
of the distribution amplitude part \cite{gdas-kk-2014}.
These decay constants are evaluated by the QCD sum rule often at $Q^2=1$ GeV$^2$,
so that they are evolved to the average scale 
$\langle Q^2 \rangle =16.6$ GeV$^2$ of the used Belle data.
The scale dependence of the decay constants indicates 
$f_{f} (Q^2 \to \infty) = 0$, which means the resonance contributions
vanish in the scaling limit. In this limit, there are sum rules for the GDAs as
\begin{align}
\int_0^1 dz \,  \Phi_q^{\,\pi^0 \pi^0} (z,\zeta,W^2) = 0, \ \ \ \ 
\int_0^1 dz \,  (2z -1) \, \Phi_q^{\,\pi^0 \pi^0} 
(z,\zeta,0) = - 4 \, M_{2(q)}^{\pi^0}  \, \zeta (1-\zeta) .
\label{eqn:gda-sum-I=0}
\end{align}
Here, $M_{2(q)}^h$ is the momentum fraction carried 
by flavor-$q$ quarks and antiquarks in the pion, so that
the total quark fraction is $\sum_q M_{2(q)}^{\pi^0}$.
The first terms of $\widetilde{B}_{10}$ and $\widetilde{B}_{12}$
in Eq.\,(\ref{eqn:B10B12-final}) are constrained at $W^2=0$
by the second sum rule.
From Eqs.\,(\ref{eqn:integral-over-z}), (\ref{eqn:emt-ffs-timelike-0}),
and (\ref{eqn:gda-parametrization}), we obtain the timelike gravitational
form factors expressed by the S- and D-wave terms of the GDAs as
\begin{align}
\Theta_{1, q} (s) 
= -\frac{3}{5} \widetilde B_{10} (W^2) + \frac{3}{10} \widetilde B_{20} (W^2) ,
\ \ \ 
\Theta_{2, q} (s) 
= \frac{9}{10 \, \beta^2} \widetilde B_{20} (W^2) ,
\label{eqn:emt-ffs-gdas}
\end{align}
and the total timelike gravitational form factors of the pion
is obtained by adding them as
$ \Theta_{n} (s) = \sum_{i=q} \Theta_{n, i} (s)$ where $n=1$ or 2.
We notice that the form factor $\Theta_{2}$ originates from the D-wave part
of the GDAs and $\Theta_{1}$ from both S- and D-wave terms.
The overall form factor for the continuum term is taken as
$ F^{\,\pi}_q (W^2) = 1 / [ 1 + (W^2-4 m_\pi^2)/\Lambda^2 ]^{n-1}$
with the cutoff $\Lambda$ and the factor $n$ taken as
the constituent-counting value $n=2$ \cite{counting}.

The $f_0 (980)$ contribution is not included in the analysis
because the differential cross-section data of the Belle
do not show such contribution
in the invariant-mass region at $W \simeq 1$ GeV and
because there is no theoretical estimate on the decay constant
by reflecting its exotic nature, tetra-quark or $K\bar K$ configuration.
There is a QCD-sum-rule calculation by assuming an ordinary $q\bar q$-type
configuration for $f_0 (980)$; however, calculated cross sections
are totally in contradiction to the Belle data. It means that
the $q\bar q$ structure is not supported by the differential cross 
section data of the Belle collaboration
for $f_0 (980)$ as it has been claimed for a long time \cite{f0-4q}.

\begin{wrapfigure}[10]{r}{0.40\textwidth}
   \vspace{-0.50cm}
   \begin{center}
     \includegraphics[width=5.0cm]{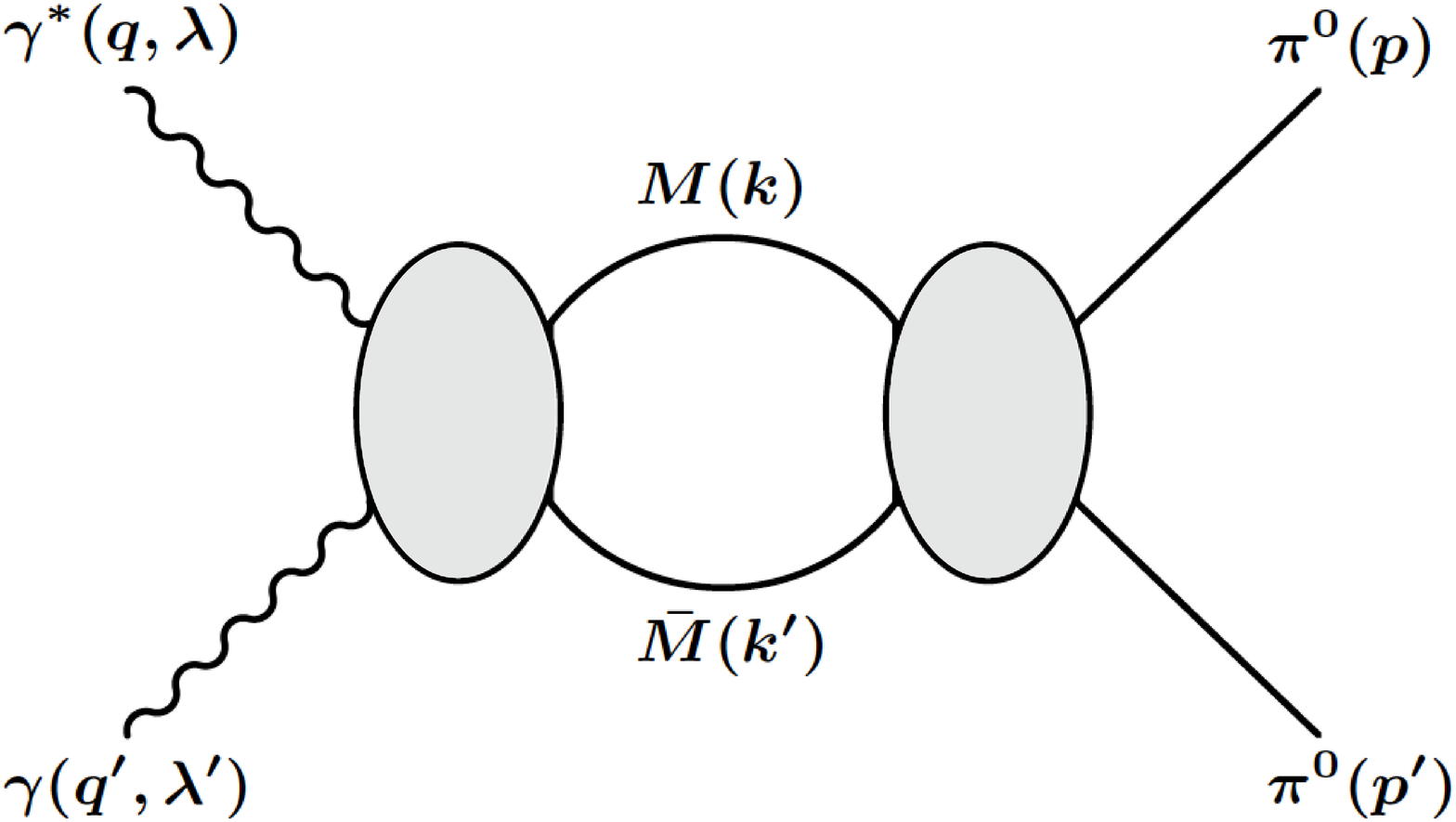}
   \end{center}
\vspace{-0.60cm}
\caption{Intermediate-meson contribution.}
\label{fig:2gamma-int-meson-2}
\vspace{-0.7cm}
\end{wrapfigure}
Some explanations are needed for understanding Eq.\,(\ref{eqn:B10B12-final}).
First, we comment on coupling constants.
There are theoretical estimates on the decay constant $f_{f_2}$,
and another one $\bar f_{f_0}$ is considered as a parameter
in our analysis because there is no theoretical estimate.
The coupling constants $g_{f_0\pi\pi}$ and $g_{f_2\pi\pi}$ are 
determined by the $2 \pi$ decay widths.
Second, there are contributions to the cross section 
$\gamma^* \gamma \to \pi^0 \pi^0$ from the processes with intermediate
mesons as shown in Fig.\,\ref{fig:2gamma-int-meson-2}.
Considering the $\pi\pi$ intermediate state, we need to include
the $\pi\pi$ phase shifts $\delta_0 (W)$ and $\delta_2 (W)$ in 
Eq.\,(\ref{eqn:B10B12-final}).
In our analysis, we use the $\pi\pi$ phase shifts 
by Bydzovsky, Kaminski, Nazari, and Surovtsev \cite{pi-pi-code}.
Above the $K\bar K$ threshold at $2 m_{K^+}=0.987$ GeV, 
the $K\bar K$ channel opens, and then $\eta\eta$ channel
also opens at higher energies. We do not include these effects
in our current analysis; however, we introduce additional
parameters in the $\pi\pi$ phase shifts above the $K\bar K$ threshold
and they are determined by fitting the Belle data.

\section{Analysis results}
\label{results}

The Belle experimental data for $\gamma^* \gamma \to \pi^0 \pi^0$
are analyzed for extracting the pion GDAs. For this purpose,
we select the data with large enough $Q^2$ to satisfy the factorization 
condition $Q^2 \gg W^2,\, \Lambda_{\text{QCD}}^2$
so that the amplitude is factorized into the hard perturbative QCD part 
and the soft GDA one. As such a condition, we take the scale
$Q^2 \ge 8.92$ GeV$^2$ for the Belle data. Then, there are 550 points 
of data with the scales $Q^2=$8.92, 10.93, 13.37, 17.23, and 24.25 GeV$^2$.
The invariant-mass range is $ 0.5 \ \text{GeV} < W < 2.1 \ \text{GeV}$,
and the scattering angles are $\cos\theta=0.1$, 0.3, 0.5, 0.7, and 0.9
in the data.

\begin{figure}[t!]
\vspace{0.10cm}
\begin{minipage}{\textwidth}
\begin{tabular}{lc}
\hspace{-0.30cm}
\begin{minipage}[c]{0.48\textwidth}
   \vspace{-0.2cm}
   \begin{center}
     \includegraphics[width=7.0cm]{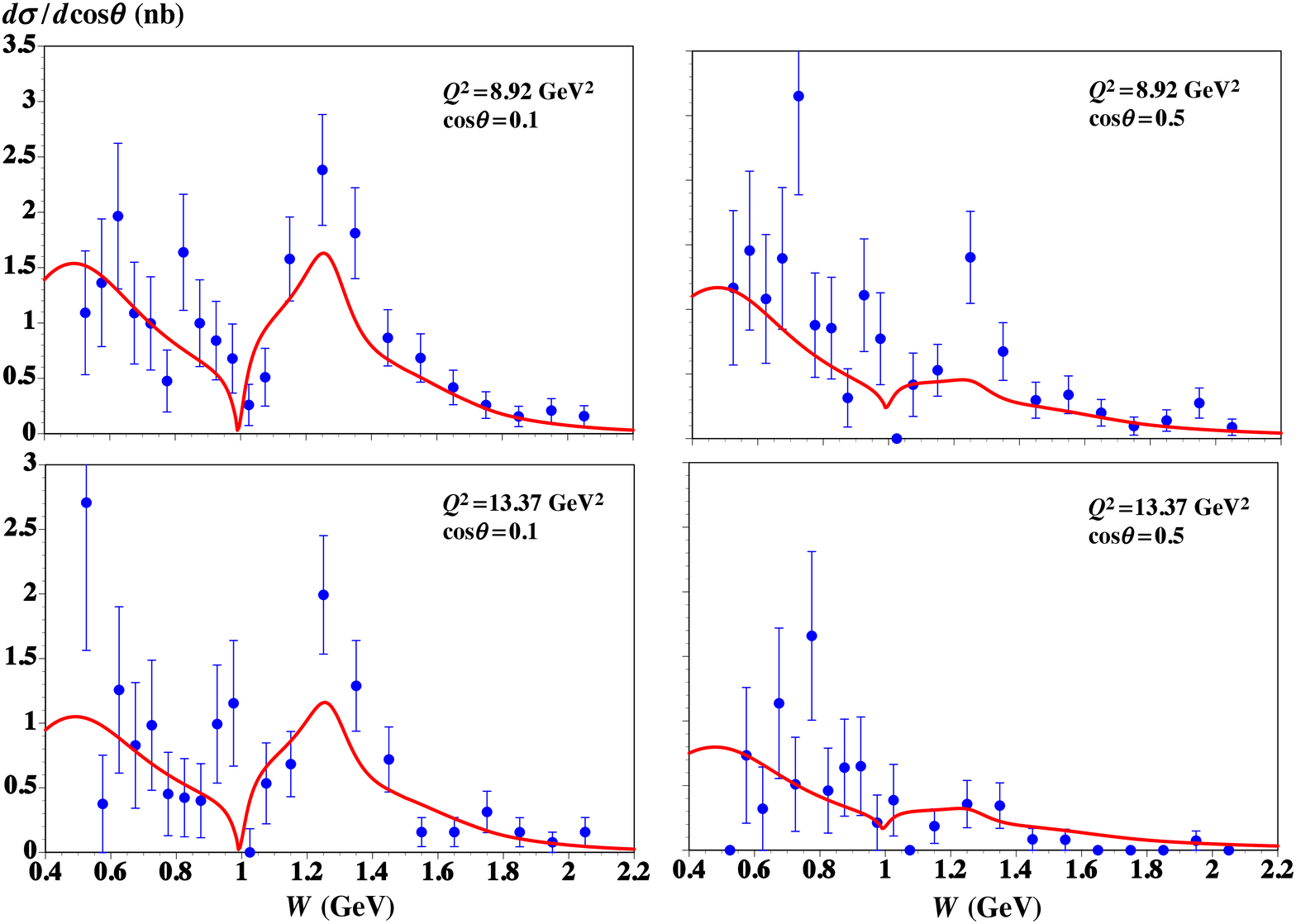}
   \end{center}
\vspace{-0.70cm}
\caption{Comparison with Belle data \cite{gdas-kst-2017}.}
\label{fig:belle-comparison}
\vspace{-0.4cm}
\end{minipage} 
\hspace{0.5cm}
\begin{minipage}[c]{0.49\textwidth}
    \vspace{-0.2cm}
   \begin{center}
    \includegraphics[width=6.0cm]{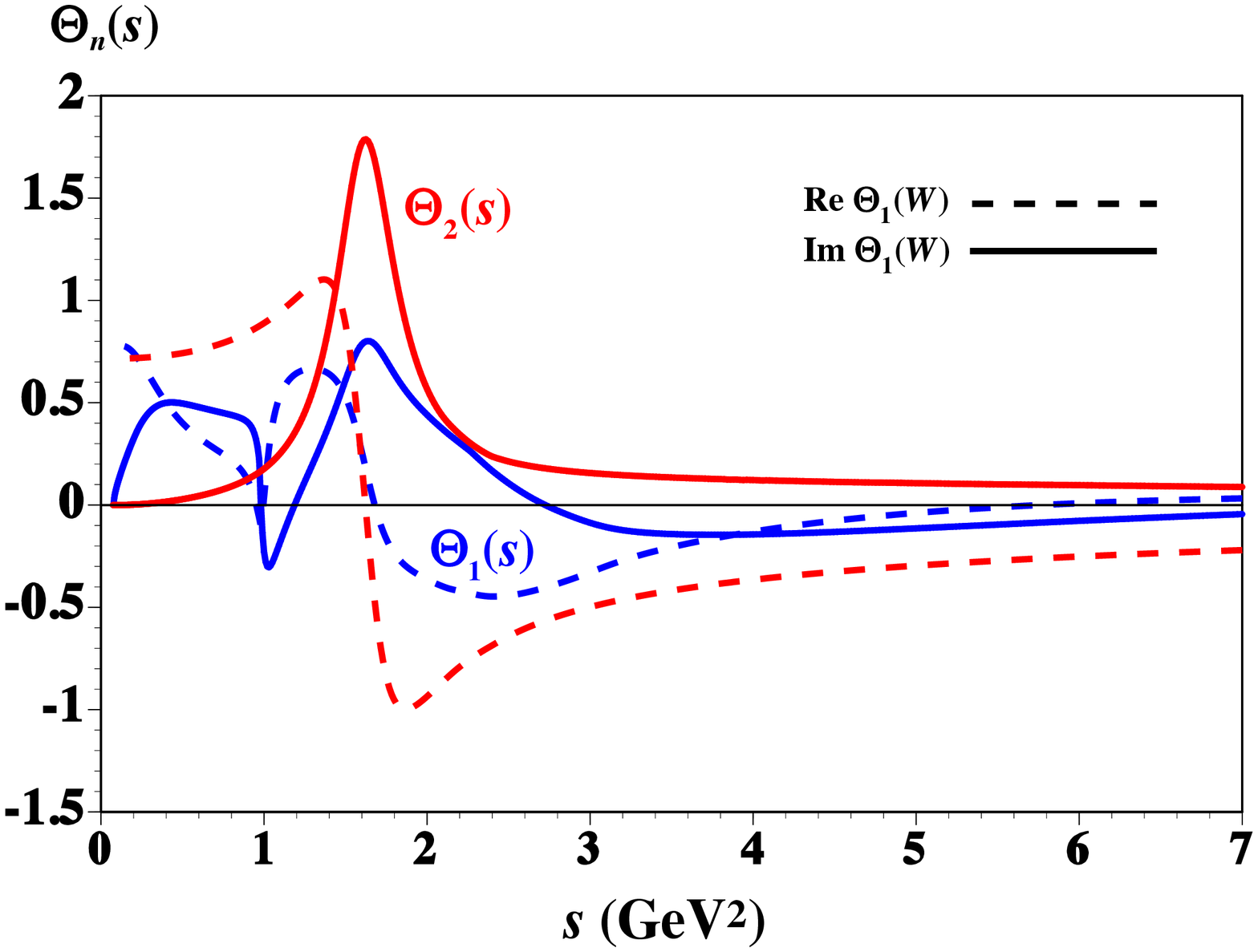}
   \end{center}
\vspace{-0.70cm}
\caption{Timelike gravitational form factors for $\pi$ \cite{gdas-kst-2017}.}
\label{fig:theta12reim-s}
\vspace{-0.4cm}   
\end{minipage}
\end{tabular}
\vspace{0.20cm}
\end{minipage}
\end{figure}

The pion GDAs are determined by fitting these cross-section data
with the equations in the previous section \cite{gdas-kst-2017}. 
The theoretical cross sections are compared with the Belle data,
as an example, at $Q^2=$8.92, 13.37 GeV$^2$ and $\cos\theta=0.1$, 0.5
in Fig.\,\ref{fig:belle-comparison}. 
Here, the results are shown by introducing phase parameters
for the S-wave phase.
We obtained a reasonable result to explain the data.
At $\cos\theta=0.1$, there is a conspicuous peak from $f_2 (1270)$
in the cross section; however, it becomes relatively small 
at $\cos\theta=0.5$. There is a $f_0 (500)$ effect at small $W$, 
and it overlaps with the continuum term in Eq.\,(\ref{eqn:B10B12-final}).
From the determined GDAs, we calculate the timelike gravitational
form factors in Fig.\,\ref{fig:theta12reim-s} by using
Eq.\,(\ref{eqn:emt-ffs-gdas}).
Since they are timelike, they contain both real and imaginary parts.
In $\Theta_2$, the $f_2 (1270)$ resonance feature is clear 
at $\sqrt{s}=1.27$ GeV, whereas $\Theta_1$ has more complicated 
$s$ dependence with both S- and D-wave contributions.

\begin{figure}[t!]
\vspace{0.10cm}
\begin{minipage}{\textwidth}
\begin{tabular}{lc}
\hspace{-0.30cm}
\begin{minipage}[c]{0.48\textwidth}
   \vspace{-0.2cm}
   \begin{center}
    \includegraphics[width=6.0cm]{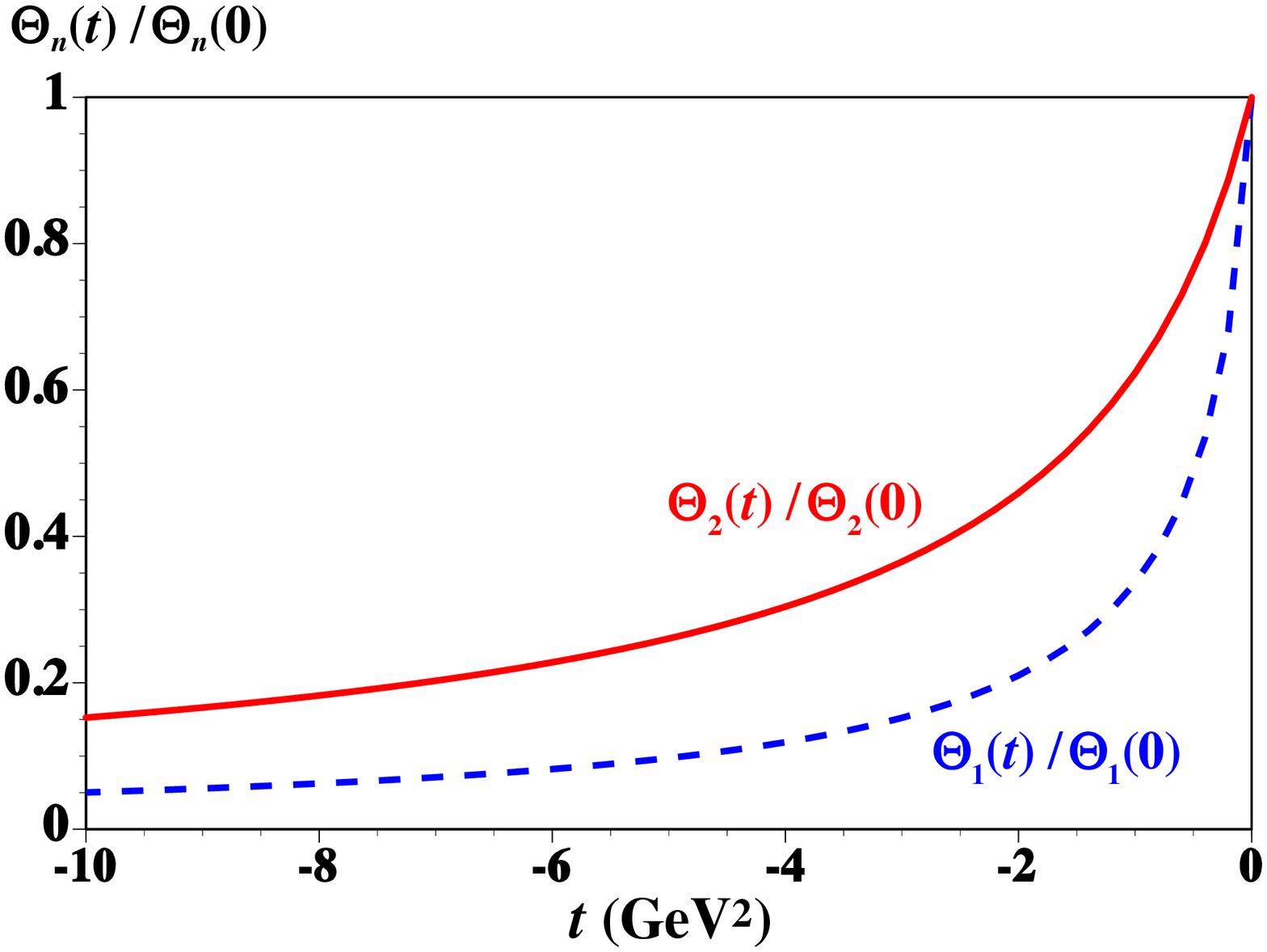}
   \end{center}
\vspace{-0.70cm}
\caption{Spacelike gravitational form factors \cite{gdas-kst-2017}.}
\label{fig:theta12-spaceline}
\vspace{-0.4cm}
\end{minipage} 
\hspace{0.5cm}
\begin{minipage}[c]{0.48\textwidth}
    \vspace{-0.2cm}
   \begin{center}
    \includegraphics[width=6.0cm]{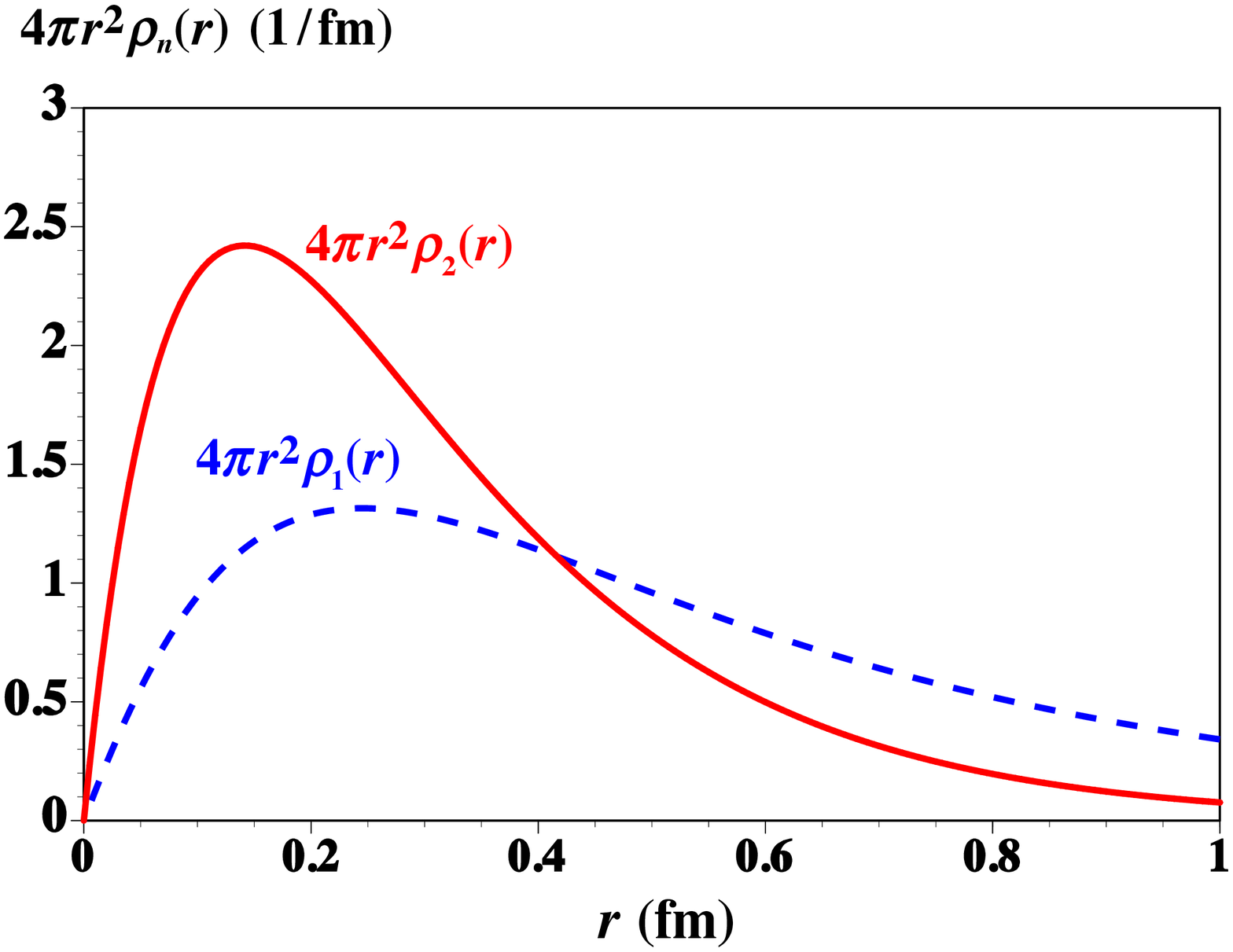}
   \end{center}
\vspace{-0.70cm}
\caption{Mass and mechanical densities \cite{gdas-kst-2017}.}
\label{fig:rho12}
\vspace{-0.4cm}   
\end{minipage}
\end{tabular}
\vspace{0.20cm}
\end{minipage}
\end{figure}

The timelike form factors are converted to the spacelike ones by
using the dispersion integral over the real positive $t$ ($\equiv s$)
with the consideration that singularities of the form factor $\Theta_n (t)$ 
($n=1,\,2$)
is in the positive real $t$ axis from $4 m_\pi^2$:
\begin{align}
\Theta_n (t) & = \int_{4 m_h^2}^\infty \frac{ds}{\pi} 
            \frac{{\rm Im}\, \Theta_n (s)}{s-t-i\varepsilon} .
\label{eqn:dispersion-form-1}
\end{align}
Then, the space-coordinate densities are calculated by
the Fourier transforms of the spacelike form factors:
\begin{align}
\rho_n (r)  = \int \frac{d^3 q}{(2\pi)^3} 
          \, e^{-i\vec q \cdot \vec r} \, \Theta_n  (q)
 = \int_{4 m_h^2}^\infty \frac{ds}{4\pi^2} 
              \frac{e^{-\sqrt{s} r}}{r} \, {\rm Im} \, \Theta_n  (s) .
\label{eqn:3D-rho}
\end{align}
Using these equations, we obtain the form factors and densities
in Figs.\,\ref{fig:theta12-spaceline} and \ref{fig:rho12} \cite{gdas-kst-2017}.

Physics meaning of these form factors and densities is understood 
in the following way. The static energy-momentum tensor may be defined 
by the three-dimensional Fourier transform as \cite{static-form} 
$ T^{\mu\nu}_q (\vec r \,) = 
\int d^3 q / [(2\pi)^3 \, 2E] e^{i \vec q \cdot \vec r} 
\left\langle \pi^0 (p') \! \left| 
T^{\mu\nu}_q (0) \, \right| \! \pi^0 (p) \right\rangle $
with the pion energy $E=\sqrt{m_\pi^2 +\vec q^{\ 2}/4}$.
The $\mu\nu = 00$ component satisfies the mass relation
$ \int d^3 r \, T^{00}_q (\vec r \,) = m_\pi \Theta_{2,q} (0)$,
which means that $\Theta_2$ and $\rho_2 (r)$ indicate
the mass (energy) distributions in the pion.
At finite $t$, $\Theta_1$ also contributes to the mass distribution.
The $\mu\nu = ij$ ($i,\, j =1,\, 2,\, 3$) components are generally
written as
$ T^{\, ij}_q (\vec r \,) = p_q (r) \, \delta_{ij} 
    + s_q (r) ( r_i r_j / r^2 - \delta_{ij} /3 )$
in terms of the pressure $p(r)$ and shear force $s(r)$.
Since $T^{\, ij}_q (\vec r \,)$ is expressed by only $\Theta_1$,
$\Theta_1$ and $\rho_1 (r)$ indicate pressure and shear-force distributions 
in the pion. We may call $\Theta_2$ and $\rho_2 (r)$ the gravitational mass 
(or energy) form factor and density, and $\Theta_1$ and $\rho_1 (r)$
may be called the mechanical (pressure, shear force) form factor and density.
As shown in Figs.\,\ref{fig:theta12-spaceline} and \ref{fig:rho12},
the mass density has a harder distribution than the mechanical one.
From the form factors or densities, we obtain 
the root-mean-square (rms) radii for both distributions:
$\sqrt {\langle r^2 \rangle _{\text{mass}}} = 0.69 \, \text{fm}$ and
$\sqrt {\langle r^2 \rangle _{\text{mech}}} = 1.45 \, \text{fm}$.
We mentioned that the parameters are introduced for the S-wave phase; 
however, an equally good fit is also obtained by assigning them to the D-wave
part. In the D-wave case, our results are slightly different
and there are some ambiguities on this assignment.
By considering this ambiguity, we obtain 
the evaluated gravitational radii as \cite{gdas-kst-2017}:
\begin{align}
\sqrt {\langle r^2 \rangle _{\text{mass}}} 
    =  0.56 \sim 0.69 \, \text{fm}, \ \ \ 
\sqrt {\langle r^2 \rangle _{\text{mech}}} 
    = 1.45 \sim 1.56 \, \text{fm} .
\label{eqn:g-radii-pion-range}
\end{align}
This is the first report on the gravitational radii 
for a hadron from experimental measurements.
It is especially interesting to find that the mass radius is similar
or slightly smaller than the charge radius
$\sqrt {\langle r^2 \rangle _{\text{charge}}} =0.672 \pm 0.008$ fm
and that the mechanical radius is larger.
However, the studies of the gravitational form factors and radii for
hadrons are still in the beginning stage, and further investigations
are needed. 

We believe that this new field has bright future to understand
gravitational physics from microscopic quark and gluon level.
Gravity studies are mainly on macroscopic systems because
gravitational interactions are ultra-weak ones and they cannot
be detected generally in microscopic particle-physics measurements.
However, as we explained in this report, we can find
the gravitational source originates from quarks and gluons by
using the technique of hadron tomography, namely by the 3D
structure functions. The KEKB will produce much accurate
cross-section data in the near future by the upgraded super-KEKB,
so that the errors in Fig.\,\ref{fig:belle-comparison} should
become much smaller in a few years.
Furthermore, the 3D structure functions can be investigated
at various high-energy facilities in the world such as 
LHC, RHIC, CERN-COMPASS, JLab, Fermilab, J-PARC, GSI, and ILC.
Time has come to investigate the 3D tomography including
the GDAs for clarifying gravitational properties of hadrons.

\section{Summary}
\label{summary}

We have determined the GDAs, gravitational form factors, and densities
for the pion by analyzing the Belle cross-section measurements
on the two-photon process $\gamma^* \gamma \to \pi^0 \pi^0$.
The GDAs are provided with several parameters by considering
the continuum term and resonances ones, and they are determined 
from the Belle data. By taking the first moments of the GDAs,
we obtained the timelike gravitational form factors 
($\Theta_1$, $\Theta_2$) for the pion.
Using the dispersion relation, they are converted to the spacelike
form factors. Then, the space-coordinate distributions 
($\rho_1 (r)$, $\rho_2 (r)$) and rms radii,
$\sqrt {\langle r^2 \rangle _{\text{mass}}} =0.56 \sim 0.69$ fm 
and $\sqrt {\langle r^2 \rangle _{\text{mech}}} =1.45 \sim 1.56$ fm,
are calculated.
The functions $\Theta_2$ and $\rho_2 (r)$ have the meaning of
the gravitational mass (energy) form factor and density, 
and $\Theta_1$ and $\rho_1 (r)$ are the mechanical (pressure, shear force) 
form factor and density.
This should be the first finding on the gravitational form factors
and radii for a hadron by analyzing actual experimental measurements.
The charge radius of the pion is $0.672 \pm 0.008$ fm.
It is our interesting finding that 
the gravitational mass radius is similar to this charge radius or
slightly smaller, and the mechanical radius is larger.

Since the 3D tomography has been a very popular topic in hadron physics
in the last several years, much progress is expected in this
novel field of gravitational physics from the fundamental
quark and gluon level. Gravitational physics in microscopic systems
had been a speculative project for a long time due to ultra-weak
interaction nature. However, time has come to investigate it 
in the microscopic level because the gravitational source from
quarks and gluons can be determined, as we showed in this work.
Our work is just the beginning of such studies,
and much progress is expected in this new research area.

\section*{Acknowledgement}
This work was supported by Japan Society for the Promotion of Science (JSPS)
Grants-in-Aid for Scientific Research (KAKENHI) Grant Number JP25105010.



\end{document}